\documentclass[aps,prb,twocolumn,superscriptaddress,showpacs]{revtex4}
\usepackage{graphicx}
\usepackage{amssymb}
\usepackage{amsmath}
\pdfoutput=1

\begin{document}

\title{Microwave response of a magnetic single-electron transistor}

\author{Scott A. Bender}
\author{Yaroslav Tserkovnyak}
\affiliation{Department of Physics and Astronomy, University of California, Los Angeles, California 90095, USA}
\author{Arne Brataas}
\affiliation{Department of Physics, Norwegian University of Science and Technology, NO-7491 Trondheim, Norway}

\date{\today}

\begin{abstract}
We consider a single-electron transistor in the form of a ferromagnetic dot in contact with normal-metal and pinned ferromagnetic leads.  Microwave-driven precession by the dot induces a pumped electric current. In open circuits, this pumping produces a measurable reverse bias voltage, which can be enhanced and made highly nonlinear by Coulomb blockade in the dot. The dependence of this bias on the power and spectrum of microwave irradiation may be utilized to develop nanoscale microwave detectors analogous to single-electron transistor-based electrostatic sensors and nanoelectromechanical devices.
\end{abstract}

\pacs{72.15.Gd,72.25.Ba,85.75.-d}


\maketitle

Recent work has demonstrated both theoretically\cite{tserkovPRB08tb,xiaoPRB08cp} and experimentally\cite{moriyamaPRL08} that a dc electric current may be pumped through a ferromagnet$\mid$insulator$\mid$ferromagnet (F$\mid$I$\mid$F) tunnel junction by pinning one ferromagnet and precessing the other at frequency $\omega$. This is analogous to spin pumping by a precessing ferromagnet into adjacent normal metals,\cite{tserkovPRL02sp} which can subsequently induce a voltage across the ferromagnet by spin-flip processes.\cite{wangPRL06vg} In these cases, the voltage generated by ferromagnetic dynamics is substantially smaller than $\hbar\omega$ (the quantum of energy supplied by the microwave source) in the absence of spin-spin or electron-electron correlation effects.

In this paper, we study the interplay of ferromagnetic pumping and Coulomb blockade in single-electron transistors, which suggests for their use as sensitive detectors of microwave irradiation. Our proposal complements and extends into the magnetic realm the established techniques utilizing single-electron transistors, such as electrostatic sensing\cite{zhitenevNAT00} and mechanical electron shuttling.\cite{scheibleAPL04}

\begin{figure}
\includegraphics[width=\linewidth,clip=]{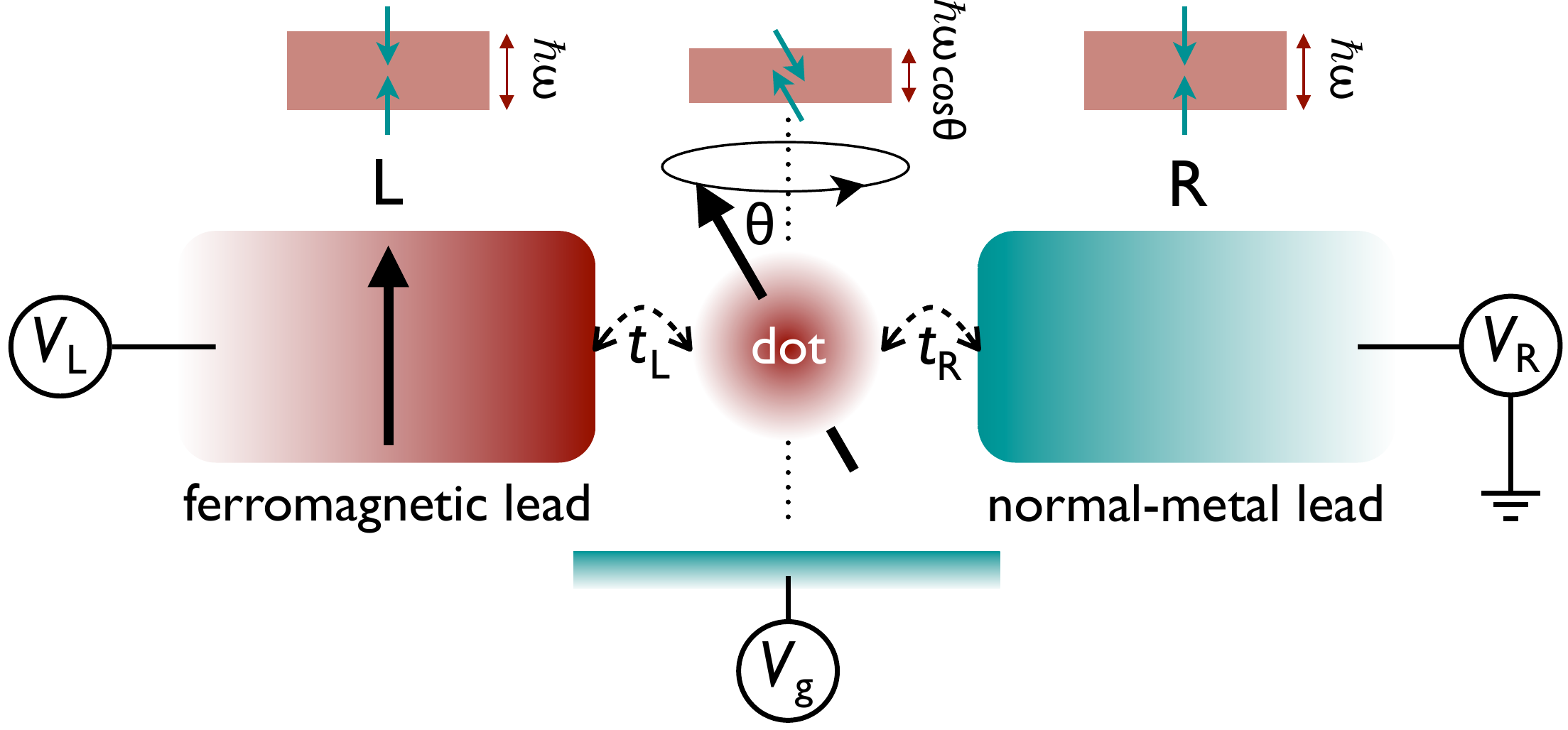}
\caption{(Color online) Schematics of the precessing magnetic dot coupled to two large reservoirs  and the effective spin splittings of the chemical potentials associated with the fictitious Zeeman field of $\hbar\omega$, according to Eqs.~(\ref{eq:new H}) and (\ref{eq:approx H}), in the rotating frame of reference. The long black arrows show magnetization directions.}
\label{sc}
\end{figure}

We consider charge pumping by a microwave-driven ferromagnetic dot with a classically large spin resonantly precessing at frequency $\omega$ (see Fig.~\ref{sc}).  The zero-dimensional nature of the quantum dot makes the electron-electron interactions relevant. Unlike static theoretical arrangements involving voltage-driven  transport between an interacting quantum dot and magnetic leads,\cite{Weymann2005yq} ours exhibits steady charge pumping by the magnetization precession. Quantum tunneling of the large magnetic moment is assumed to be strongly suppressed by the dissipative environment of the phonon continuum and/or electronic excitations associated with metallic regions; the dynamics of the dot are therefore dominated by classical precession, in contrast to the proposed macroscopic quantum tunneling of the dot's magnetic moment in Ref.~\onlinecite{haiNAT09}. Traditional parametric spin and charge pumping by varying tunneling amplitudes and energy-level structure\cite{brouwerPRB98} in a strongly-interacting normal quantum dot contacted by magnetic reservoirs was considered recently in Refs.~\onlinecite{splettstoesserPRB08}. 

In open circuits, the charge pumping induces an electrostatic buildup between the right and left leads, which we represent here as the bias $V_{0}\equiv V_R-V_L$ that yields $I=0$. (We will henceforth consider the right reservoir to be grounded, i.e., $V_R=0$.)  Without Coulomb blockade, this bias is linear in pumping frequency:\cite{tserkovPRB08tb}
\begin{equation}
V_{0}=\frac{\hbar\omega}{2e}\frac{P\sin^{2}\theta}{1+P^{2}\cos\theta}\,,
\label{eq: FIF yaroslav's}
\end{equation}
where $P=(D_\uparrow-D_\downarrow)/(D_\uparrow+D_\downarrow)$ is the polarization of the dot and ferromagnetic lead in terms of the spin-dependent density of states $D_s$ ($-e$ is the electron charge). The nominal \textit{charge-pumping efficiency} $\mathcal{E} \equiv eV_{0}/\hbar\omega$ [as well as the differential efficiency $\mathcal{E}_{\mathrm{diff}}\equiv(e/\hbar)\partial V_0/\partial \omega$] is independent of the microwave frequency and small, vanishing as $\theta^2$ when the precession angle goes to zero. To be specific, the dot is taken to be made of the same material as the ferromagnetic lead.

We demonstrate here that electron-electron interaction on the dot gives rise to a highly nonlinear response $V_{0}(\omega)$, which is also robust at small $\theta$. The efficiency $\mathcal{E}$ of this response is greatly increased (although still less than one), while the differential efficiency $\mathcal{E}_{\mathrm{diff}}$ can become extremely large when $\hbar\omega$ is close to the Coulomb-blockade energy gap. This frequency (or, equivalently, Coulomb gap) sensitivity of $ \mathcal{E}_{\mathrm{diff}}$ may pave way for microwave spectral analyzer and magnetoelectronic logic applications.

Central to our discussion is the observation that the precessing dot creates a fictitious spin-dependent voltage; this bias, in turn, drives electron transport via hopping onto and off of the dot from two metallic leads, one nonmagnetic (``right'' lead) and one with spin-exchange splitting $\Delta$ in the $z$ direction (``left" lead) (cf. Fig.~\ref{sc}). Supposing the dot is steadily precessing clockwise around the $z$ axis at a constant angle $\theta$, the total single-electron Hamiltonian (without including electron interactions on the dot) can be written as $\hat{H}(t)=\mathbf{p}^{2}/2m+V(\mathbf{r})-\Delta\mathbf{m}(\mathbf{r},t)\cdot\hat{\boldsymbol{\sigma}}/2$, where $\hat{\boldsymbol{\sigma}}$ is a vector of Pauli matrices and $\mathbf{m}$ is the majority spin direction. The first two terms determine the tunneling Hamiltonian and energy spectra of the leads and dot, while $\mathbf{m}(\mathbf{r},t)$ is given by $(0,0,1)$ in the left lead, $(\sin\theta\cos(\omega t),\sin\theta\sin(\omega t),\cos\theta)$ in the dot, and is set to zero everywhere else.  
 By going into the rotating frame of reference, the precession of the dot is formally eliminated, at the expense of transforming the Hamiltonian as follows:\cite{tserkovPRB05}
\begin{equation}
\hat{H}(t)\to \hat{R}^{\dagger}\hat{H}\hat{R}-i\hbar\hat{R}^{\dagger}\partial_{t}\hat{R}=\hat{H}(t=0)-(\hbar\omega/2)\hat{\sigma}_z,\,
\label{eq:new H}
\end{equation}
where $\hat{R}=e^{-i\omega t\hat{\sigma}_z/2}$ is a rotation operator that transforms out dot precession while leaving the spin-independent energy terms (including Coulomb interaction) unaffected. Whereas, according to Eq.~(\ref{eq:new H}), the lead Hamiltonians pick up a fictitious spin-dependent potential $-(\hbar\omega/2)\hat{\sigma}_z$, the dot Hamiltonian can be simplified in the rotating frame to
\begin{equation}
\hat{H}_{\rm dot}(t)\to \hat{H}_{\rm dot}(0)-(\hbar\omega/2)\hat{\sigma}_\parallel\cos\theta\,,
\label{eq:approx H}
\end{equation}
where $\hat{\sigma}_\parallel=\hat{\boldsymbol{\sigma}}\cdot\mathbf{m}_{\rm dot}(0)$ is the spin operator in the direction of the $t=0$ dot magnetization $\mathbf{m}_{\rm dot}$, and we have disregarded the normal component of the fictitious field in the dot, which is valid in conventional ferromagnets with $\hbar\omega\ll\Delta$. The extra ``inertia" terms in Eqs.~(\ref{eq:new H}) and (\ref{eq:approx H}) shift energies of the spin-up (down) electrons by $\mp\hbar\omega/2$ in the leads and $\mp(\hbar\omega/2)\cos\theta$ in the dot, thus creating an effective spin-dependent bias between leads and dot that can drive  transport currents. Assuming strong spin relaxation in the dot, on the scale of the electron injection rate, no spin accumulation is built up there.

In the sequential tunneling regime, the electric current flowing from, say, the left (ferromagnetic) lead to the metallic dot is given by a sum over the possible number of electrons $N$ occupying the dot:\cite{nazarovBOOK09} $I_L=-e\sum_{N}P(N)\left(\Gamma_{N\rightarrow N+1}^L-\Gamma_{N\rightarrow N-1}^L\right)$,
where $\Gamma_{N\rightarrow N\pm1}^L$ is the tunneling rate for one electron to hop from (to) the ferromagnet to (from) the $N$-occupied dot and $P(N)$ is the probability that $N$ excess electrons are contained on the dot at a given moment of time.  Coulomb blockade effects are captured by introducing the electrostatic energy $E_{N}$ associated with $N$ electrons occupying the dot, where $E_N=E_cN(N-1)/2-eV_gN$
and $V_g$ is the gate voltage (renormalized by various mutual capacitances). The energy for adding a single electron to the $N$-electron dot is $\mu_N\equiv E_{N+1}-E_N=E_cN-eV_g$.
Setting the equilibrium chemical potential of the leads to zero, the dot operates at the characteristic electron number $N\sim eV_g/E_c$. The energy scale $E_{c}$ is realistically of the order of 10~meV, while the driving energy $\hbar\omega$ is typically not more than a fraction of an meV. This requires going to Kelvin-range temperatures if one is to completely disregard thermal effects. We suppose the gate voltage $V_{g}$ on the dot can be tuned so that the gap for adding one excess electron is within the range of the driving frequency $\hbar\omega$, but higher occupancies are increasingly less likely due to a finite $E_c$.

 Let us discuss a sufficiently large $E_{c}$, such that only the transitions $N\rightleftharpoons N+1$ between the dot and the leads are relevant. The dot electrons occupy parallel and antiparallel spin states that adiabatically evolve with the precessing magnetization. Both leads are also considered to be equilibrated in the lab frame of reference at the respective spin-independent voltages, so the tunneling rate for each is a sum over four channels for two spin projections of dot electrons hopping to (from) static up and down states in the leads:
\begin{align}
\Gamma_{N\rightleftharpoons N+1}^{(l)}   =&k_BT\left|t_l\right|^{2}\sum_{s,s'}D_{s}^{(l)}D_{s'}\left|\langle s|s'\rangle_\theta\right|^{2}\nonumber\\
\times& f\left(\pm\left[\mu_N+eV_l+(\hbar\omega/2)(s-s'\cos\theta)\right]\right).
\label{eq:f plus current}
\end{align}
Here, $l=L,R$ labels the left/right leads and $s=\uparrow, \downarrow$ (or $\pm$), spin projection along the magnetization direction (or $z$ axis for the normal lead), while $D_{s}^{(L)}=D_{s}$, $D_{s}^{(R)}=D$ are the ferromagnetic and normal-metal densities of states,  respectively. We consider the tunnel amplitudes $t_L$ and $t_R$ to be energy independent. The spin-space matrix elements squared are given by: $\left|\langle\uparrow|\uparrow\rangle_\theta\right|^{2}=\left|\langle\downarrow|\downarrow\rangle_\theta\right|^{2}=\cos^{2}(\theta/2)$ and $\left|\langle\uparrow|\downarrow\rangle_\theta\right|^{2}=\sin^{2}(\theta/2)$. The temperature-dependent weighting function in Eq.~(\ref{eq:f plus current}) is $f(\epsilon)=(\epsilon/k_{B}T)(e^{\epsilon/k_{B}T}-1)^{-1}$.

\begin{figure}
\includegraphics[width=\linewidth,clip=]{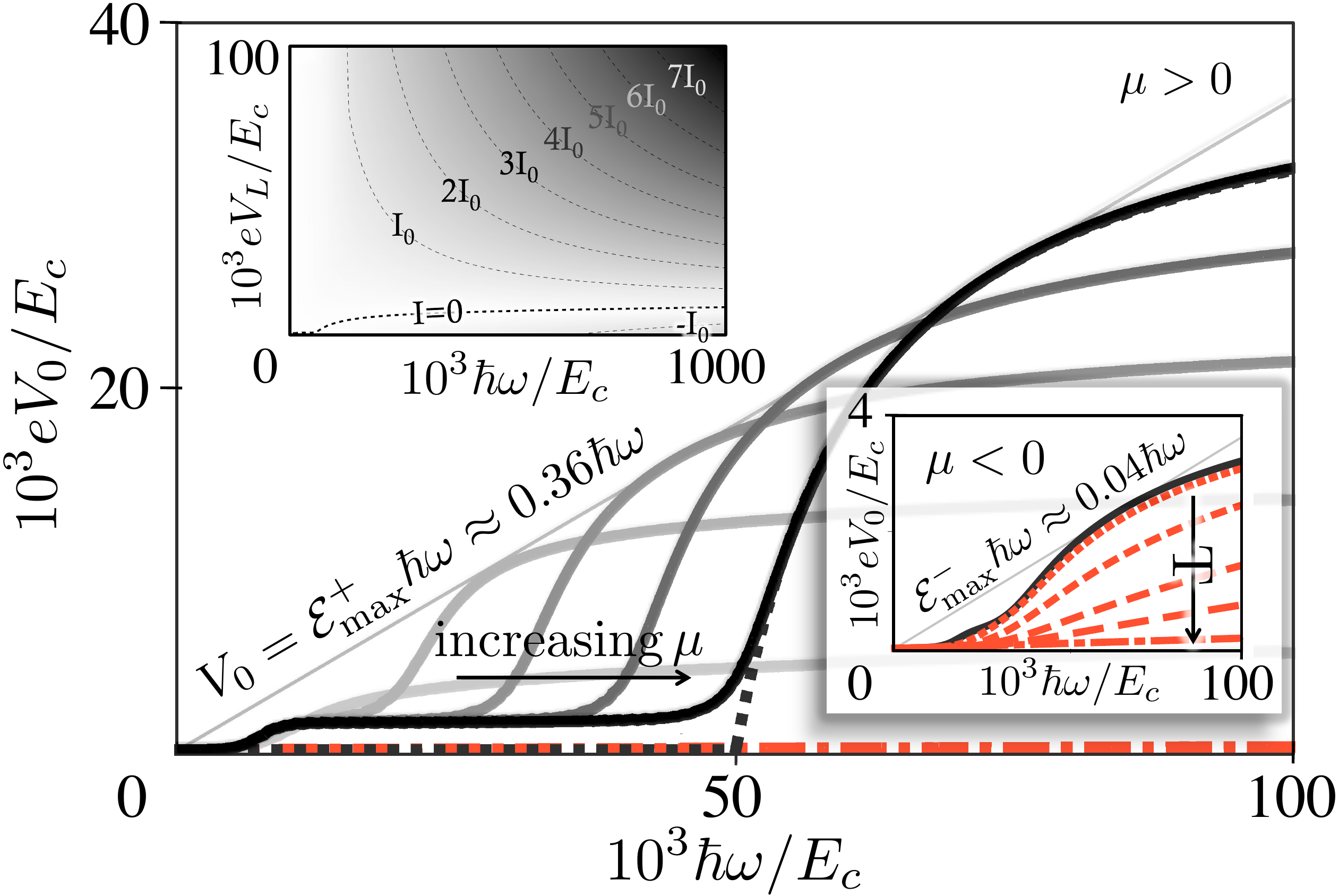}
\caption{(Color online) Low-frequency $I=0$  ($\hbar\omega\ll E_{c}$) numerical curves for $\mu>0$, $\theta=5^{\circ}$, $P=2/3$, and $k_{B}T/E_{c}=10^{-3}$. Here, the increasingly darker gray lines represent $\mu/E_{c}=(1,2,3,4,5)\times10^{-2}$, respectively, while the dotted-dashed red line corresponds to zero Coulomb blockade, Eq.~(\ref{eq: FIF yaroslav's}). The $T=0$ small-angle analytic solution, Eq.~(\ref{eq:pos G}),  is shown as a dotted black line superimposed on the corresponding finite-temperature curve in black. Upper inset: High-frequency relief plot of current density for the same parameters as the solid black curve in the main panel, with $t_L=t_R=t$ and  $I_0=0.02\,ek_BTD^2 |t|^2$. Lower inset: thermal effects for $\mu/E_{c}=-3\times10^{-2}$. The black curve corresponds to $k_{B}T/E_{c}=10^{-3}$ and the increasingly long red dashed lines to $k_{B}T/E_{c}=(5,6,7,8)\times10^{-3}/2$, respectively. The dotted-dashed red line illustrates the zero Coulomb-blockade case, as in the main panel.}
\label{posMu}
\end{figure}

Let us count $N$ with respect to a reference state, such that for $(k_BT,\hbar\omega)\ll E_{c}$ the dot switches between $N=0$ and $N=1$ occupancies, henceforth denoting $\mu\equiv E_1-E_0$. The total steady-state electric current, $I=I_L=I_R$, is then:
\begin{equation}
I=-e\frac{\Gamma_{0\rightarrow1}^L\Gamma_{1\rightarrow0}^R-\Gamma_{1\rightarrow0}^L\Gamma_{0\rightarrow1}^R}{\Gamma_{0\rightarrow1}^L+\Gamma_{0\rightarrow1}^R+\Gamma_{1\rightarrow0}^L+\Gamma_{1\rightarrow0}^R}\,.
\label{eq: Current}
\end{equation}
The current as a function of $V_{L}$ (with $V_R=0$) and $\omega$ is graphed in an inset of Fig.~\ref{posMu}. Under the transformation $ \mu_{N}\rightarrow-\mu_{N}$, $V_R\rightarrow-V_R$, $V_L\rightarrow-V_L$, and $\omega\rightarrow-\omega$, the electric current (\ref{eq: Current}) changes sign,  reflecting the electron-hole symmetry in our model. We can therefore choose to consider only positive $\omega$. According to Eq.~(\ref{eq: Current}), the condition for zero current is
\begin{equation}
\Gamma_{0\rightarrow1}^L\Gamma_{1\rightarrow0}^R=\Gamma_{1\rightarrow0}^L\Gamma_{0\rightarrow1}^R\,.
\label{eq:i=0 condition}
\end{equation}
The microwave-induced potential $V_{0}$ is thus independent of $D$, $t_L$, or $t_R$ and depends only on $\omega$, $P$, $\mu$, and $T$.

\begin{figure}
\includegraphics[width=\linewidth]{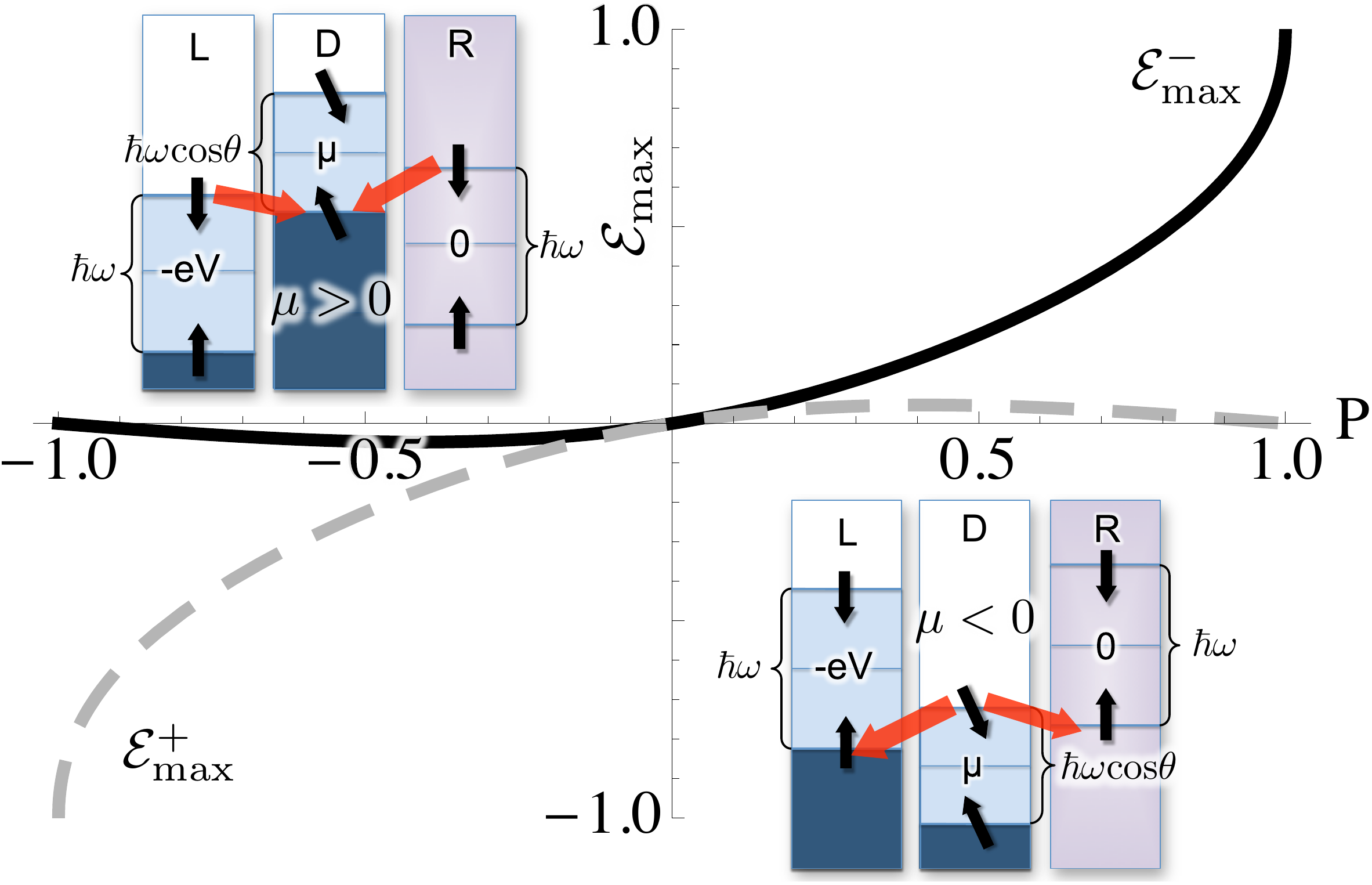}
\caption{(Color online) Maximum efficiencies $\mathcal{E}_{\rm max}^{+}$ and $\mathcal{E}_{\rm max}^{-}$ for positive and negative gating $\mu$, respectively, at zero temperature and small angles $\theta$, obtained from Eq.~(\ref{eq:pos G}). $\mathcal{E}_{\rm max}^+(P)=-\mathcal{E}_{\rm max}^-(-P)$. The inset schematics illustrate the difference between the two cases. The short black arrows show the effective spin-up/down chemical potentials in the dot and the leads. $\mu\gtrless0$ corresponds to the empty/occupied dot ($N=0/1$) in equilibrium, which becomes populated/emptied by spin-flipped tunneling (shown by the long red arrows towards/from the dot) when $\hbar\omega>|\mu|$.}
\label{maxE}
\end{figure}

Let us now turn to the zero-temperature properties, which can be found analytically. First, at arbitrary $\mu$ in the limit $\theta\to\pi/2$,
\begin{equation}
eV_{0}=P\frac{\mu^2-(\hbar\omega/2)^{2}}{P\mu-\hbar\omega/2}\Theta\left(\hbar\omega/2-\left|\mu\right|\right),
\label{eq: ninety degrees}
\end{equation}
where the Heaviside step function $\Theta\left(x\right)$ reflects the fact that the Coulomb-blockaded transport is blocked at low frequencies. At zero gap ($\mu=0$), we have $V_{0}=P\hbar\omega/2e$, in accord with Eq.~(\ref{eq: FIF yaroslav's}). At nonzero gap, the system exhibits marginally increased or decreased charge pumping efficiency $\mathcal{E}$ over the noninteracting value (\ref{eq: FIF yaroslav's}), depending on $\mu$ and $P$.  Second, we consider the limit $\theta\rightarrow0$, keeping $\mu$ finite. Eq.~(\ref{eq: FIF yaroslav's}) mandates that the charge pumping vanishes with $\theta\to0$ when the electron-electron interactions are neglected. At a nonzero Coulomb-blockade gap $\mu$, however, the induced voltage remains finite as $\theta\rightarrow0$ and, in fact, is dramatically enhanced compared to the $\theta=\pi/2$ result, Eq.~(\ref{eq: ninety degrees}). At exactly $\theta=0$, the total current vanishes, as it should, and $V_{0}=0$. However, as $\theta\rightarrow0$, we obtain angle-independent zero-temperature solutions for $\mu\gtrless0$ given by
\begin{equation}
eV_{0}=\frac{P\mu(\mu\mp\hbar\omega)}{P\mu-[(1+P^2)/(1\pm P)]\hbar\omega}\Theta\left(\hbar\omega\mp\mu\right)\,.
\label{eq:pos G}
\end{equation}
Again, the step function $\Theta \left( x\right)$ shows the current to be blocked for small frequencies up to $\hbar \omega=\left| \mu \right|$, where the response $V_0$ abruptly switches on.
It should be clear, however, that the limit $(\theta,T)\to0$ is nonanalytic: a finite $T$ makes $V_0$ vanish in the limit of $\theta\to0$, progressively more abruptly so at small temperatures. The physical explanation for a finite $V_0$ at small angles and low temperatures in the presence of Coulomb blockade is as follows. 
First, we need to appreciate the importance of hopping involving a spin flip, although the rates of these processes with respect to the equilibrium state vanish as $\theta^2$ (i.e., the spin-flip matrix elements squared) according to Eq.~(\ref{eq:f plus current}).
The insets of Fig.~\ref{maxE}  show as long red arrows the slower, bottlenecking step of the two-part sequential process for transport of a charge from lead to dot to opposite lead, in the presence of Coulomb blockade. The voltage $V_0$ would then develop in response to this weak out-of-equilibrium tunneling. The backaction of the voltage on tunneling will not be appreciable, however, until it approaches a finite value on the scale set by energies $\hbar\omega-|\mu|$ and $\mu$, leading to Eq.~(\ref{eq:pos G}). Note that, based on this reasoning, we should anticipate that the time necessary for the build-up of a finite voltage (\ref{eq:pos G}) diverges as $\theta\to0$, since the spin-flipped pumping rates vanish as $\theta^2$.

\begin{figure}
\includegraphics[width=0.9\linewidth,clip=]{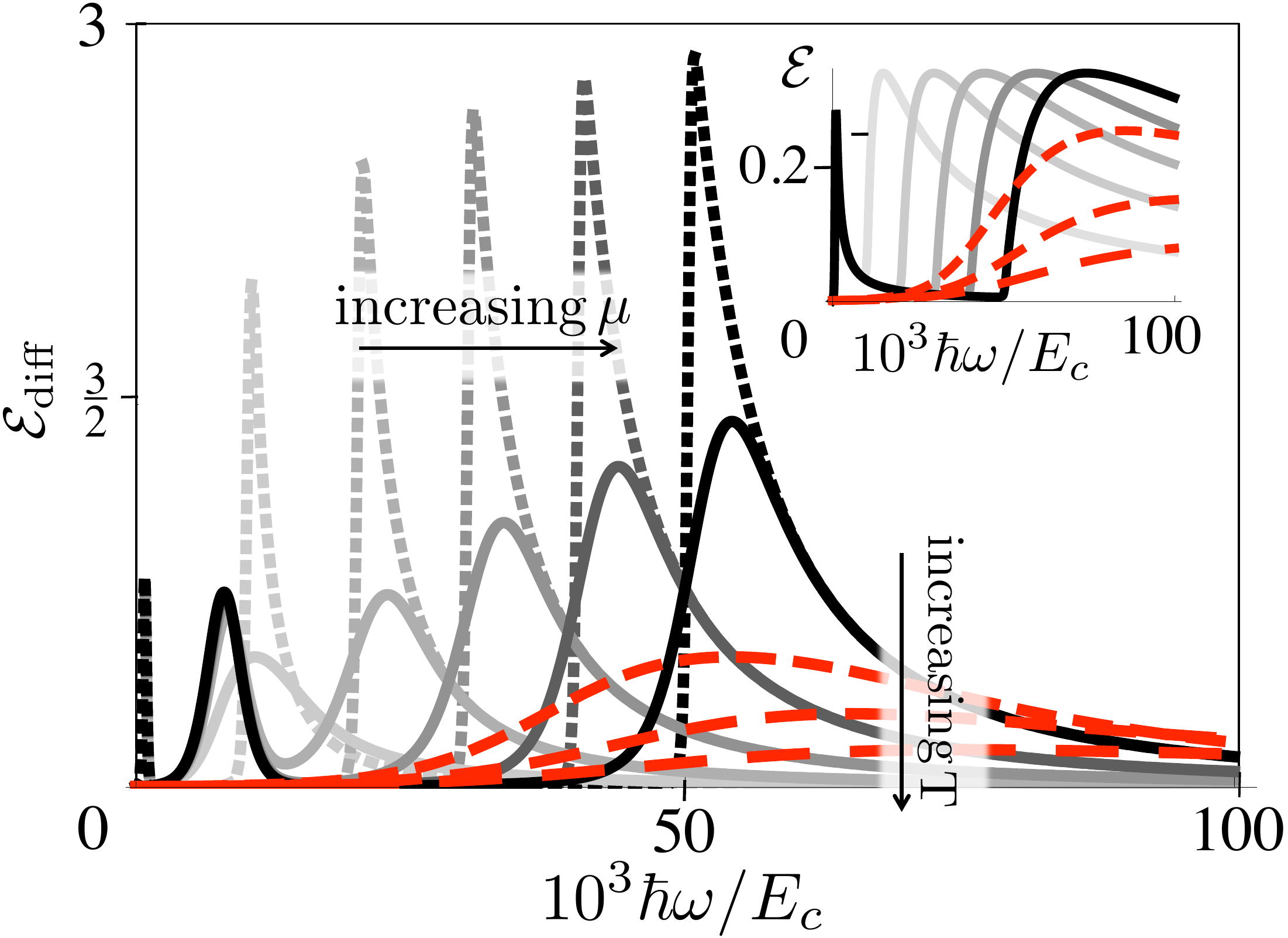}
\caption{(Color online) The solid grayscale curves in the main panel show the differential charge-pumping efficiency $\mathcal{E}_{\mathrm {diff}}=\left(e/\hbar\right) \partial V_0/\partial \omega$ for $\mu/E_{c}=(1,2,3,4,5)\times10^{-2}$ at $\theta=5^{\circ}$, $P=2/3$, and $k_{B}T/E_{c}=10^{-3}$, according to Eq.~(\ref{eq:closed}). The dotted lines show sharper efficiency peaks as the temperature is lowered to $k_{B}T/E_{c}=10^{-4}$. The dashed red lines show smearing of the peaks as the temperature is increased to $k_{B}T/E_{c}=(5,6,7)\times10^{-3}$ for $\mu/E_{c}=5\times10^{-2}$. Note that $\mu=0$ efficiency is too small to be seen. Inset: The nominal charge-pumping efficiency $\mathcal{E}=eV_0/\hbar\omega$ for the same parameters (omitting the $k_{B}T/E_{c}=10^{-4}$ data).}
\label{Eff}
\end{figure}

At finite temperatures and arbitrary $\mu$ and $\theta$, Eq.~(\ref{eq:i=0 condition}) is transcendental in $V_{0}$ and must be solved numerically or approximately. When the induced voltage is low, expanding Eq.~(\ref{eq:i=0 condition}) in  $V_0$ gives 
\begin{equation}
V_{0}\left(\omega\right)\approx\left.\frac{\Gamma_{0\rightarrow1}^{R}\Gamma_{1\rightarrow0}^{L}-\Gamma_{1\rightarrow0}^{R}\Gamma_{0\rightarrow1}^{L}}{\Gamma_{1\rightarrow0}^{R}\partial_{V_L}\Gamma_{0\rightarrow1}^{L}-\Gamma_{0\rightarrow1}^{R}\partial_{V_L}\Gamma_{1\rightarrow0}^{L}}\right|_{V_L,V_R=0},
\label{eq:closed}
\end{equation}
which can be used to find the pumping efficiencies $\mathcal{E}$ and $\mathcal{E}_{\mathrm {diff}}$ (see Fig.~\ref{Eff}). We have numerically graphed $V_{0}$ versus $\omega$ for various positive $\mu$ at $k_{B}T/E_{c}=10^{-3}$ and $\theta=5^{\circ}$ in Fig.~\ref{posMu}, and confirmed that the analytical curves obtained from Eq.~(\ref{eq:closed}) (not shown) reproduce the numerical ones very closely. At low frequencies, the response $V_{0}$ is linear in $\omega$, due to thermal excitations. At higher frequencies, the response increases gradually before plateauing at some $V_{0}$, the value of which depends on the sign of the gap $\mu$. For both signs of $\mu$, the plateau sets in at about $\hbar\omega/E_c\sim10^{-2}$. However, once $\hbar\omega$ reaches $\left|\mu\right|$, the microwave driving starts to take over the Coulomb blockade, and $V_{0}$ increases rapidly (see, e.g., the dotted line in Fig.~\ref{posMu} for zero temperature). At some $\omega$, this increase begins to fall off and, at high enough frequencies, the response becomes linear and of essentially the same slope as $\mu=0$, albeit with an offset. 

It can be noticed from Fig.~\ref{posMu} (see also the inset in Fig.~\ref{Eff}) that $\mathcal{E}(\omega)$ attains some maximum value $\mathcal{E}_{\rm max}$ that depends only on the sign of $\mu$ (and the ferromagnetic polarization $P$) at low temperatures. We can straightforwardly obtain these $\mathcal{E}^\pm_{\rm max}(P)$ ($\pm$ here labeling positive/negative $\mu$, respectively) from the zero-temperature expression, Eq.~(\ref{eq:pos G}), valid at small $\theta$. See Fig.~\ref{maxE} for the corresponding plots. For the parameters in Fig.~\ref{posMu}, $\mathcal{E}^{+}_{\rm max}\approx0.36$ and $\mathcal{E}^{-}_{\rm max}/\mathcal{E}^{+}_{\rm max}\approx0.1$, while the noninteracting efficiency (\ref{eq: FIF yaroslav's}) is only $\mathcal{E}\approx0.0018$.

The reason for different maximum efficiencies for opposite $\mu$ can be understood as follows. For a dot attractive to one electron (i.e., $\mu<0$), the bottleneck process in sequential tunneling is releasing the electron off the dot, i.e., $\Gamma^{L,R}_{1\rightarrow 0}$. Just above the threshold frequency $\hbar\omega=|\mu|$, the only contributing process to these rates is from electrons that spin-flip from a down-state in the dot to an up-state in the reservoirs (see the upper inset of Fig.~\ref{maxE}).  \textit{Both} of these processes are proportional to the number of available spin-down states in the dot, $D_\downarrow$. In contrast, for a dot repulsive to an extra electron (i.e., $\mu>0$), at the same threshold pumping, the bottleneck processes $\Gamma^{L,R}_{0\rightarrow 1}$ represent electrons tunneling from a down-state in the leads to an up-state in the dot (see the lower inset of Fig.~\ref{maxE}), both being proportional to the number of available up-states in the dot, $D_\uparrow$. One should notice, furthermore, that in the $\mu>0$ case, these bottleneck channels for tunneling into the two leads (which are here supplying the majority-spin electrons for the dot) become progressively more asymmetric between the two leads as $P\to1$. We can, therefore, expect greater absolute maximum efficiency $|\mathcal{E}^{+}_{\rm max}|$ for a $\mu>0$ dot when $P>0$ and a greater absolute maximum efficiency $|\mathcal{E}^{-}_{\rm max}|$ for a $\mu<0$ dot when  $P<0$, which is exactly what we find for the $\theta\rightarrow 0$ case in Fig.~\ref{maxE}. In fact, $\mathcal{E}^\pm_{\rm max}\to1$ and $0$, respectively, as $P\to 1$. By the aforementioned electron-hole symmetry, $\mathcal{E}^\pm_{\rm max}$ switch roles when $\omega\to-\omega$, which corresponds to a different circular polarization of the ferromagnetic precession.

Finally, supposing one has a coherent source of microwaves of unknown wavelength, the microwave frequency can be measured by ensuring that the dot is in resonance with the source and slowly ramping the electrostatic gate voltage from $\mu=E_{c}/2$ down to zero, until the onset of strong charge pumping at $\mu=\hbar\omega\cos^2({\theta}/2)$ (supposing $\theta<\pi/2$, to be specific).  While this would require a frequency less than $E_{c}/2\hbar\cos^2({\theta}/2)$ (or else other transitions become relevant) and low temperatures, it should be simple to detect the dramatic onset of pumping, either by the reverse bias $V_0$ directly or the differential efficiency $\mathcal{E}_{\mathrm{diff}}$. Further, we note that by gating the dot so that it is occupied by one electron, we have a single-electron transistor, as evidenced by the zero-frequency $I$-$V$ characteristics. By instead gating with the pumping frequency $\hbar\omega$, we can achieve an extremely high differential voltage gain $\mathcal{E}_{\mathrm{diff}}=(e/\hbar)\partial V_0/\partial \omega$ at the onset of nonzero response $V_{0}$ (see Fig.~\ref{Eff}). This offers a potential for the on-chip integration of such devices with highly-tunable and coherent microwave sources provided by the nanomagnet spin-torque oscillators.\cite{kiselevNAT03}


This work was supported in part by the Alfred~P. Sloan Foundation, DARPA, and the NSF under Grant No. DMR-0840965.


\begin{thebibliography}{17}
\expandafter\ifx\csname natexlab\endcsname\relax\def\natexlab#1{#1}\fi
\expandafter\ifx\csname bibnamefont\endcsname\relax
  \def\bibnamefont#1{#1}\fi
\expandafter\ifx\csname bibfnamefont\endcsname\relax
  \def\bibfnamefont#1{#1}\fi
\expandafter\ifx\csname citenamefont\endcsname\relax
  \def\citenamefont#1{#1}\fi
\expandafter\ifx\csname url\endcsname\relax
  \def\url#1{\texttt{#1}}\fi
\expandafter\ifx\csname urlprefix\endcsname\relax\def\urlprefix{URL }\fi
\providecommand{\bibinfo}[2]{#2}
\providecommand{\eprint}[2][]{\url{#2}}

\bibitem[{\citenamefont{Tserkovnyak et~al.}(2008)\citenamefont{Tserkovnyak,
  Moriyama, and Xiao}}]{tserkovPRB08tb}
\bibinfo{author}{\bibfnamefont{Y.}~\bibnamefont{Tserkovnyak}},
  \bibinfo{author}{\bibfnamefont{T.}~\bibnamefont{Moriyama}}, \bibnamefont{and}
  \bibinfo{author}{\bibfnamefont{J.~Q.} \bibnamefont{Xiao}},
  \bibinfo{journal}{Phys. Rev. B} \textbf{\bibinfo{volume}{78}},
  \bibinfo{eid}{020401(R)} (\bibinfo{year}{2008}).

\bibitem[{\citenamefont{Xiao et~al.}(2008)\citenamefont{Xiao, Bauer, and
  Brataas}}]{xiaoPRB08cp}
\bibinfo{author}{\bibfnamefont{J.}~\bibnamefont{Xiao}},
  \bibinfo{author}{\bibfnamefont{G.~E.~W.} \bibnamefont{Bauer}},
  \bibnamefont{and} \bibinfo{author}{\bibfnamefont{A.}~\bibnamefont{Brataas}},
  \bibinfo{journal}{Phys. Rev. B} \textbf{\bibinfo{volume}{77}},
  \bibinfo{eid}{180407} (\bibinfo{year}{2008}).

\bibitem[{\citenamefont{Moriyama et~al.}(2008)\citenamefont{Moriyama, Cao, Fan,
  Xuan, Nikoli{\'c}, Tserkovnyak, Kolodzey, and Xiao}}]{moriyamaPRL08}
\bibinfo{author}{\bibfnamefont{T.}~\bibnamefont{Moriyama}},
  \bibinfo{author}{\bibfnamefont{R.}~\bibnamefont{Cao}},
  \bibinfo{author}{\bibfnamefont{X.}~\bibnamefont{Fan}},
  \bibinfo{author}{\bibfnamefont{G.}~\bibnamefont{Xuan}},
  \bibinfo{author}{\bibfnamefont{B.~K.} \bibnamefont{Nikoli{\'c}}},
  \bibinfo{author}{\bibfnamefont{Y.}~\bibnamefont{Tserkovnyak}},
  \bibinfo{author}{\bibfnamefont{J.}~\bibnamefont{Kolodzey}}, \bibnamefont{and}
  \bibinfo{author}{\bibfnamefont{J.~Q.} \bibnamefont{Xiao}},
  \bibinfo{journal}{Phys. Rev. Lett.} \textbf{\bibinfo{volume}{100}},
  \bibinfo{eid}{067602} (\bibinfo{year}{2008}).

\bibitem[{\citenamefont{Tserkovnyak et~al.}(2002)\citenamefont{Tserkovnyak,
  Brataas, and Bauer}}]{tserkovPRL02sp}
\bibinfo{author}{\bibfnamefont{Y.}~\bibnamefont{Tserkovnyak}},
  \bibinfo{author}{\bibfnamefont{A.}~\bibnamefont{Brataas}}, \bibnamefont{and}
  \bibinfo{author}{\bibfnamefont{G.~E.~W.} \bibnamefont{Bauer}},
  \bibinfo{journal}{Phys. Rev. Lett.} \textbf{\bibinfo{volume}{88}},
  \bibinfo{eid}{117601} (\bibinfo{year}{2002});
\bibinfo{author}{\bibfnamefont{Y.}~\bibnamefont{Tserkovnyak}},
  \bibinfo{author}{\bibfnamefont{A.}~\bibnamefont{Brataas}},
  \bibinfo{author}{\bibfnamefont{G.~E.~W.} \bibnamefont{Bauer}},
  \bibnamefont{and} \bibinfo{author}{\bibfnamefont{B.~I.}
  \bibnamefont{Halperin}}, \bibinfo{journal}{Rev. Mod. Phys.}
  \textbf{\bibinfo{volume}{77}}, \bibinfo{eid}{1375} (\bibinfo{year}{2005}).

\bibitem[{\citenamefont{Wang et~al.}(2006)\citenamefont{Wang, Bauer, van Wees,
  Brataas, and Tserkovnyak}}]{wangPRL06vg}
\bibinfo{author}{\bibfnamefont{X.}~\bibnamefont{Wang}},
  \bibinfo{author}{\bibfnamefont{G.~E.~W.} \bibnamefont{Bauer}},
  \bibinfo{author}{\bibfnamefont{B.~J.} \bibnamefont{van Wees}},
  \bibinfo{author}{\bibfnamefont{A.}~\bibnamefont{Brataas}}, \bibnamefont{and}
  \bibinfo{author}{\bibfnamefont{Y.}~\bibnamefont{Tserkovnyak}},
  \bibinfo{journal}{Phys. Rev. Lett.} \textbf{\bibinfo{volume}{97}},
  \bibinfo{eid}{216602} (\bibinfo{year}{2006});
\bibinfo{author}{\bibfnamefont{M.~V.} \bibnamefont{Costache}},
  \bibinfo{author}{\bibfnamefont{M.}~\bibnamefont{Sladkov}},
  \bibinfo{author}{\bibfnamefont{S.~M.} \bibnamefont{Watts}},
  \bibinfo{author}{\bibfnamefont{C.~H.} \bibnamefont{van~der Wal}},
  \bibnamefont{and} \bibinfo{author}{\bibfnamefont{B.~J.} \bibnamefont{van
  Wees}}, \textit{ibid.} \textbf{\bibinfo{volume}{97}},
  \bibinfo{eid}{216603} (\bibinfo{year}{2006}).

\bibitem[{\citenamefont{Zhitenev et~al.}(2000)\citenamefont{Zhitenev, Fulton,
  Yacoby, Hess, Pfeiffer, and West}}]{zhitenevNAT00}
\bibinfo{author}{\bibfnamefont{N.~B.} \bibnamefont{Zhitenev}},
  \bibinfo{author}{\bibfnamefont{T.~A.} \bibnamefont{Fulton}},
  \bibinfo{author}{\bibfnamefont{A.}~\bibnamefont{Yacoby}},
  \bibinfo{author}{\bibfnamefont{H.~F.} \bibnamefont{Hess}},
  \bibinfo{author}{\bibfnamefont{L.~N.} \bibnamefont{Pfeiffer}},
  \bibnamefont{and} \bibinfo{author}{\bibfnamefont{K.~W.} \bibnamefont{West}},
  \bibinfo{journal}{Nature} \textbf{\bibinfo{volume}{404}},
  \bibinfo{pages}{473} (\bibinfo{year}{2000});
\bibinfo{author}{\bibfnamefont{J.}~\bibnamefont{Martin}},
  \bibinfo{author}{\bibfnamefont{S.}~\bibnamefont{Ilani}},
  \bibinfo{author}{\bibfnamefont{B.}~\bibnamefont{Verdene}},
  \bibinfo{author}{\bibfnamefont{J.}~\bibnamefont{Smet}},
  \bibinfo{author}{\bibfnamefont{V.}~\bibnamefont{Umansky}},
  \bibinfo{author}{\bibfnamefont{D.}~\bibnamefont{Mahalu}},
  \bibinfo{author}{\bibfnamefont{D.}~\bibnamefont{Schuh}},
  \bibinfo{author}{\bibfnamefont{G.}~\bibnamefont{Abstreiter}},
  \bibnamefont{and} \bibinfo{author}{\bibfnamefont{A.}~\bibnamefont{Yacoby}},
  \bibinfo{journal}{Science} \textbf{\bibinfo{volume}{305}},
  \bibinfo{pages}{980} (\bibinfo{year}{2004}).

\bibitem[{\citenamefont{Scheible and Blick}(2004)}]{scheibleAPL04}
\bibinfo{author}{\bibfnamefont{D.~V.} \bibnamefont{Scheible}} \bibnamefont{and}
  \bibinfo{author}{\bibfnamefont{R.~H.} \bibnamefont{Blick}},
  \bibinfo{journal}{Appl. Phys. Lett.} \textbf{\bibinfo{volume}{84}},
  \bibinfo{pages}{4632} (\bibinfo{year}{2004}).

  \bibitem[{\citenamefont{Konig and Martinek}(2003)\citenamefont{Konig and Martinek}}]{Weymann2005yq}
\bibinfo{author}{\bibfnamefont{J.}~\bibnamefont{K\"onig}}
 \bibnamefont{and} \bibinfo{author}{\bibfnamefont{J.}~\bibnamefont{Martinek}},
  \bibinfo{journal}{Phys. Rev. Lett.} \textbf{\bibinfo{volume}{90}},
  \bibinfo{eid}{166602} (\bibinfo{year}{2003});
\bibinfo{author}{\bibfnamefont{M.}~\bibnamefont{Braun}},
\bibinfo{author}{\bibfnamefont{J.}~\bibnamefont{K\"onig}}
 \bibnamefont{and} \bibinfo{author}{\bibfnamefont{J.}~\bibnamefont{Martinek}},
  \bibinfo{journal}{Phys. Rev. B} \textbf{\bibinfo{volume}{70}},
  \bibinfo{eid}{195345} (\bibinfo{year}{2004});
\bibinfo{author}{\bibfnamefont{I.}~\bibnamefont{Weymann}},
  \bibinfo{author}{\bibfnamefont{J.}~\bibnamefont{K\"onig}},
  \bibinfo{author}{\bibfnamefont{J.}~\bibnamefont{Martinek}},
  \bibinfo{author}{\bibfnamefont{J.}~\bibnamefont{Barna{\'s}}},  \bibnamefont{and} \bibinfo{author}{\bibfnamefont{G.}~\bibnamefont{Sch{\"o}n}},
  \textit{ibid.} \textbf{\bibinfo{volume}{72}},
  \bibinfo{pages}{115334} (\bibinfo{year}{2005});
\bibinfo{author}{\bibfnamefont{F.~M.} \bibnamefont{Souza}},
  \bibinfo{author}{\bibfnamefont{J.~C.} \bibnamefont{Egues}},
   \bibnamefont{and} \bibinfo{author}{\bibfnamefont{A.~P.} \bibnamefont{Jauho}},
  \textit{ibid.} \textbf{\bibinfo{volume}{75}},
  \bibinfo{eid}{165303} (\bibinfo{year}{2007}).
  
\bibitem[{\citenamefont{Hai et~al.}(2009)\citenamefont{Hai, Ohya, Tanaka,
  Barnes, and Maekawa}}]{haiNAT09}
\bibinfo{author}{\bibfnamefont{P.~N.} \bibnamefont{Hai}},
  \bibinfo{author}{\bibfnamefont{S.}~\bibnamefont{Ohya}},
  \bibinfo{author}{\bibfnamefont{M.}~\bibnamefont{Tanaka}},
  \bibinfo{author}{\bibfnamefont{S.~E.} \bibnamefont{Barnes}},
  \bibnamefont{and} \bibinfo{author}{\bibfnamefont{S.}~\bibnamefont{Maekawa}},
  \bibinfo{journal}{Nature} \textbf{\bibinfo{volume}{458}},
  \bibinfo{pages}{489} (\bibinfo{year}{2009}).

\bibitem[{\citenamefont{Brouwer}(1998)}]{brouwerPRB98}
\bibinfo{author}{\bibfnamefont{P.~W.} \bibnamefont{Brouwer}},
  \bibinfo{journal}{Phys. Rev. B} \textbf{\bibinfo{volume}{58}},
  \bibinfo{eid}{R10135} (\bibinfo{year}{1998}).
 
   \bibitem[{\citenamefont{Splettstoesser
  et~al.}(2008)\citenamefont{Splettstoesser, Governale, and
  K{\"o}nig}}]{splettstoesserPRB08}
\bibinfo{author}{\bibfnamefont{J.}~\bibnamefont{Splettstoesser}},
  \bibinfo{author}{\bibfnamefont{M.}~\bibnamefont{Governale}},
  \bibnamefont{and}
  \bibinfo{author}{\bibfnamefont{J.}~\bibnamefont{K{\"o}nig}},
  \bibinfo{journal}{Phys. Rev. B} \textbf{\bibinfo{volume}{77}},
  \bibinfo{eid}{195320} (\bibinfo{year}{2008});
\bibinfo{author}{\bibfnamefont{F.}~\bibnamefont{Cavaliere}}, 
\bibinfo{author}{\bibfnamefont{M.}~\bibnamefont{Governale}}, \bibnamefont{and}
  \bibinfo{author}{\bibfnamefont{J.}~\bibnamefont{K\"onig}},
  \bibinfo{journal}{Phys. Rev. Lett} \textbf{\bibinfo{volume}{103}},
  \bibinfo{eid}{136801} (\bibinfo{year}{2009}).
 
  \bibitem[{\citenamefont{Tserkovnyak and Brataas}(2005)}]{tserkovPRB05}
\bibinfo{author}{\bibfnamefont{Y.}~\bibnamefont{Tserkovnyak}} \bibnamefont{and}
  \bibinfo{author}{\bibfnamefont{A.}~\bibnamefont{Brataas}},
  \bibinfo{journal}{Phys. Rev. B} \textbf{\bibinfo{volume}{71}},
  \bibinfo{eid}{052406} (\bibinfo{year}{2005}).

\bibitem[{\citenamefont{Nazarov and Blanter}(2009)}]{nazarovBOOK09}
\bibinfo{author}{\bibfnamefont{Y.~V.} \bibnamefont{Nazarov}} \bibnamefont{and}
  \bibinfo{author}{\bibfnamefont{Y.~M.} \bibnamefont{Blanter}},
  \emph{\bibinfo{title}{Quantum Transport}} (\bibinfo{publisher}{Cambridge
  University Press}, \bibinfo{address}{Cambridge}, \bibinfo{year}{2009}).
  
\bibitem{kiselevNAT03}
S.~I. Kiselev, J.~C. Sankey, I.~N. Krivorotov, N.~C. Emley, R.~J. Schoelkopf,  R.~A. Buhrman, and D.~C. Ralph,
\newblock{Nature} {\bf 425}, 380 (2003).
 
\end{thebibliography}
\end{document}